\documentclass{article}

\usepackage{arxiv}
\usepackage{float}
\usepackage[utf8]{inputenc} 
\usepackage[T1]{fontenc}    
\usepackage{hyperref}       
\usepackage{url}            
\usepackage{booktabs}       
\usepackage{amsfonts}       
\usepackage{amsmath}
\usepackage[bibliography=common]{apxproof}
\usepackage{mathtools}
\usepackage{nicefrac}       
\usepackage{microtype}      
\usepackage{lipsum}		
\usepackage{graphicx}
\usepackage{natbib}
\usepackage{doi}
\usepackage{xcolor}
\usepackage[linesnumbered,boxed,noend]{algorithm2e}
\usepackage{booktabs}

\newtheoremrep{lemma}{Lemma}

\definecolor{electricviolet}{rgb}{0.56, 0.0, 1.0}
\definecolor{applegreen}{rgb}{0.55, 0.71, 0.0}

\title{Pangenome Peeling}

\date{} 

\author{ \href{https://orcid.org/0000-0002-3545-039X}{\includegraphics[scale=0.06]{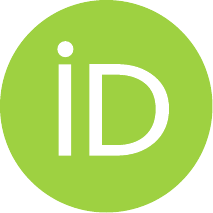}\hspace{1mm} Alitzel L\'opez S\'anchez}\\
	Department of Mathematics and Statistics\\
	University of Ottawa\\
        Ottawa, ON, Canada\\
	\texttt{alopezsa@uottawa.ca} \\
	\And
	\href{https://orcid.org/0000-0002-9390-8426}{\includegraphics[scale=0.06]{orcid.pdf}\hspace{1mm}Jiazhen Leng} \\
	Department of Mathematics and Statistics\\
	University of Ottawa\\
        Ottawa, ON, Canada\\
	\texttt{jleng100@uottawa.ca} \\
    \And
    Natasha Ivanova \\
	Department of Mathematics and Statistics\\
	University of Ottawa\\
        Ottawa, ON, Canada\\
	\texttt{nivan018@uottawa.ca} \\
	\And
    \href{https://orcid.org/0000-0002-4586-2347}{\includegraphics[scale=0.06]{orcid.pdf}\hspace{1mm}Lingling Jin} \\
	Department of Computer Science\\
	University of Saskatchewan\\
        Saskatoon, SK, Canada\\
	\texttt{lingling.jin@cs.usask.ca} \\
    \And
    \href{https://orcid.org/0000-0001-8415-5189s}{\includegraphics[scale=0.06]{orcid.pdf}\hspace{1mm}David Sankoff} \\
	Department of Mathematics and Statistics\\
	University of Ottawa\\
        Ottawa, ON, Canada\\
	\texttt{sankoff@uottawa.ca}
}

\renewcommand{\undertitle}{Decomposing Weighted Adjacencies into Path-Forest Genomes}

\renewcommand{\shorttitle}{ Pangenome Structure Optimization}

\hypersetup{
pdftitle={Pangenome Peeling: Decomposing Weighted Adjacencies into Path Forest Genomes},
pdfsubject={q-bio.NC, q-bio.QM},
pdfauthor={Alitzel L\'opez S\'anchez, Lingling Jin, Jiazhen Leng, David Sankoff},
pdfkeywords={Pangenomics, Graph Decomposition, Combinatorial Optimization, Comparative Genomics},
}

\begin{document}

\maketitle

\begin{abstract}
One aspect of pangenome analysis that has received little attention is the recovery of the individual genomes that gave rise to a pangenome graph. While substantial effort has focused on constructing pangenomes from collections of genomes, the inverse problem remains largely unexplored: given a pangenome represented as a set of weighted adjacencies, can we reconstruct the constituent linear genomes that underlie it?

Thus, we introduce the pangenome peeling problem, which seeks to decompose a weighted adjacency graph into a collection of linear genomes whose combined adjacencies reproduce the observed edge weights. This problem is particularly relevant when considering ancestral pangenomes, where the graph structure may be known or inferred, but the underlying genomes are not.

As a first step toward addressing this problem, we formally define pangenome peeling and derive a lower bound on the minimum number of genomes required to explain a given weighted adjacency graph. We then present two reconstruction methods: one based on the sequential addition of edges and a second based on extracting maximum-spanning-tree structures. Both methods produce genome collections that exactly explain the input adjacency data.

To evaluate the recovered sets of genomes, we compare them to the set of input genomes using a maximum-weight matching framework based on the double-cut-and-join distance with indels. Here, the maximum-spanning-tree approach is superior, producing sets of constituent genomes close to the simulations while the sequential addition of edges approach yields constituent genomes increasingly divergent from the input as genome sizes increase.
\end{abstract}


\section{Introduction}

The widespread adoption of long-read sequencing technologies has led to an unprecedented increase in the number of complete genomes available across diverse taxa. As genomic collections continue to expand, it has become elucidated that a single reference genome cannot capture the full extent of genetic variation within a species. To address this limitation, comparative genomic studies increasingly rely on pangenomes, which provide a unified representation of the genetic content observed across multiple individuals~\cite{jayakodi,secomandi2025pangenome}. Pangenomes have emerged as a versatile framework for studying sequence diversity, genome evolution, and population-level variation across a wide range of organisms~\cite{liu}.

The utility of pangenomic approaches has been demonstrated across numerous biological systems. Early studies revealed substantial genetic variation that remained invisible under a single-reference paradigm. Notably, the soybean pangenome uncovered genes associated with disease resistance, flowering time, and seed composition that were absent from the cultivated reference genome~\cite{li2014novo}. Since then, pangenome analyses have become central to studies of adaptation, biodiversity, and crop improvement~\cite{wang2023pangenome}.

Despite these advances, most computational research on pangenomes has focused on the forward problem: constructing a pangenome representation from a collection of genomes. In contrast, little attention has been devoted to the inverse problem. Given a pangenome representation, what information can be recovered about the genomes that underlie it? Addressing this question is essential for understanding the expressive power of pangenome models, the information they preserve, and the ambiguities they inevitably contain.

Recovering constituent genomes from aggregate pangenomic information may prove valuable in settings where complete sampling is impossible. For many lineages, extant genomes provide only a partial view of the past genetic diversity, and extinct or unsampled lineages may have contributed substantially to the observed genomic landscape. This perspective has been useful in other areas of comparative genomics, including studies of horizontal gene transfer involving extinct and unsampled taxa~\cite{lgtdead}. Recovering constituent genomes from a pangenome may provide insight into genomic structures that are no longer directly observable.

\paragraph{Our contributions.} In this work, we introduce the \emph{Pangenome Peeling Problem}, a novel inverse formulation of pangenome reconstruction. We consider a weighted adjacency graph in which vertices represent genes and edge weights record the multiplicity of observed adjacencies across a collection of genomes. While such graphs are commonly used as pangenome representations, they aggregate information from multiple genomes and therefore obscure the identity of their constituent genomes. We study the problem of recovering a collection of genomes whose combined adjacency content exactly reproduces a given weighted adjacency graph.

We formally define this reconstruction problem and derive an approximate lower bound on the minimum number of constituent genomes required to explain a given instance. We further propose two algorithms based on sequential edge addition (SEQ) and maximum-spanning-tree path decompositions (MST). Using simulated benchmarks, we evaluate the inferred genome collections with respect to pangenome size, core set recovery, and structural similarity measured by the DCJ-indel distance. Our results revealing the superiority of the MST approach.

The open-source implementation of the proposed algorithms, the pangenome simulator, and all scripts used in the experimental evaluation are available at \url{https://github.com/AliLopSan/pangesim}.

\section{Definitions and Model Specification}
Unless stated otherwise, all graphs in this work are undirected and loopless. For a graph $G$, $V(G)$ and $E(G)$ denote the sets of vertices and edges of $G$, respectively. The degree of a vertex $v$ is denoted by $deg_G(v)$ and its neighborhood by $N_G(v)$. A graph is connected if every pair of vertices is joined by a path. A connected component is a maximal connected subgraph. A path $p$ is a graph with a sequence of vertices $(v_1,v_2,\dots,v_k)$ such that $(v_i, v_{i+1}) \in E(p)$ for all $i \in \{1, \dots, k-1\}$. A path forest (also known as linear forest) is a graph whose connected components are paths. A path is non-trivial if it has at least one edge. An edge $(u,v)$ is an endpoint of $p$ if $deg_p(v) \leq 1$ or $deg_p(u) \leq 1$.

\subsection{Pangenomes and Weighted Adjacency Graphs}

Let $\mathcal{G}=\{1,\ldots,n\}$ denote a set of genomic markers. Throughout this work, markers (or genes) represent homologous sequence segments and are treated as unique identifiers.  Following the genome rearrangement literature, we model a genome as a collection of linear chromosomes and therefore represent it as a \emph{path forest} with an additional requirement. More formally, a genome is a graph $G$ where $G \subseteq \mathcal{G}$ such that every vertex has a degree at most two and every connected component is a non-trivial path. We refer to $|V(G)|$ as the size of a genome.

A \emph{pangenome} is a collection of genomes $P = \{G_1, G_2, \dots , G_k\}$ where $\bigcup_{i=1}^{k} V(G_i) = \mathcal{G}$. We refer to $k$ as the size of the pangenome. The \emph{core} of $P$ is the set of markers present in every genome $C(P) = \bigcap^k_{i=1} V(G_i)$ whereas the \emph{shell} of $P$ is the set of markers that are present in at least one genome but not in all of them. For a pangenome $P$, the \emph{endogenous weight} of an adjacency $(u,v)$ is the number of constituent genomes of $P$ which contain $(u,v)$ as an edge, that is $w_P(u,v) = |\{G \in P: (u,v) \in E(G)\}|$. Whenever the pangenome is clear from the context, we simply write $w(u,v)$.

Several graph-based pangenome representations have been proposed, including sequence graphs, genome graphs and variation graphs~\cite{eizenga,bao}. Depending on the representation, vertices may correspond to $k$-mers, genomic segments or other sequence units, while edges encode relationships such as adjacency, concatenation or variation. Paths in these graphs represent haplotypes, contigs or genomic annotations~\cite{odgi}. Although these models differ in their representation of genomic variation, they all encode adjacency relationships between genomic segments.

In this work we abstract this information into a weighted adjacency graph 
$H = (\mathcal{G}, A, m)$ where $m:A \to \mathbb{Z}_{>0}$ is the \emph{multiplicity function} which records how often an adjacency $(u,v) \in A$ was seen in the source data. 

We consider the reverse problem of reconstructing a collection of linear genomes from the aggregate adjacency information. Given a weighted adjacency graph $H=(\mathcal{G},A,m)$, we seek a pangenome whose endogenous edge weights reproduce the observed multiplicities. We say that such a pangenome $P$ \emph{explains} $H$.

\medskip

\noindent  
The {\sf Pangenome Peeling} problem.\\
\textbf{Input.} A weighted simple graph $H =(\mathcal{G}, A, m)$, where $m:A \to \mathbb{Z}_{>0}$ is the \emph{multiplicity function} which records how often an adjacency $(u,v) \in A$ was seen in the source data. \\
\textbf{Output.} A pangenome $P$ that explains $H$.
\vspace{2mm}

\begin{figure}
    \centering
    \includegraphics[width=\textwidth]{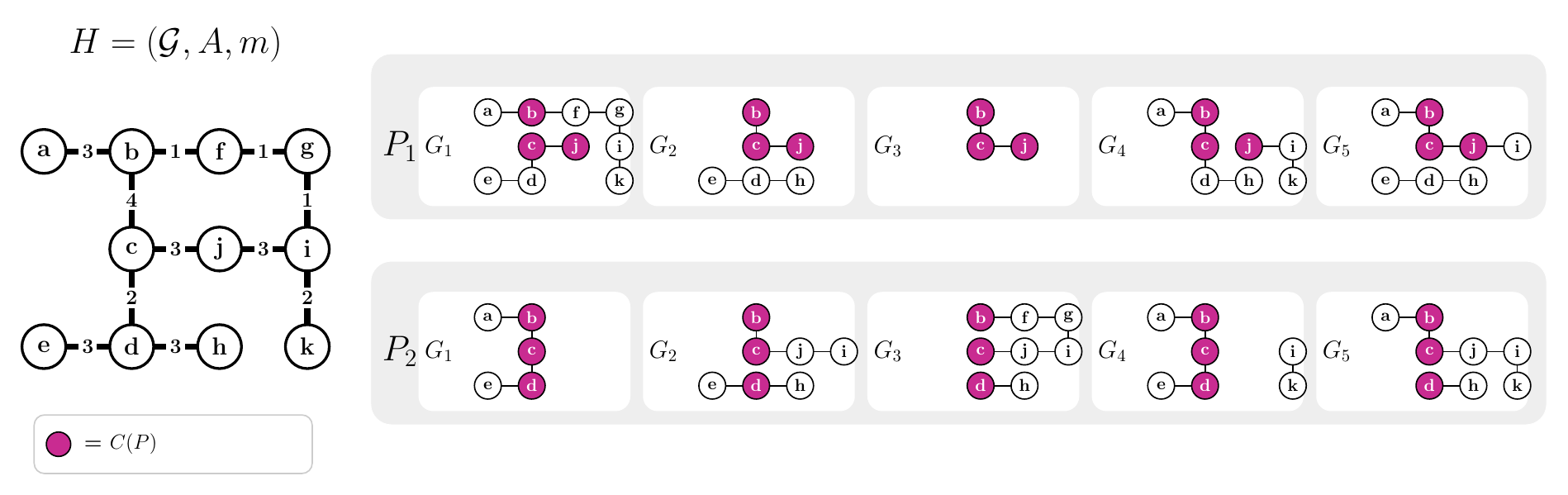}
    \caption{Given the weighted adjacency graph $H$. Pangenomes $P_1$ and $P_2$ are alternative peelings of $H$. The coloured nodes represent elements of their respective core sets.}
    \label{fig:manypan}
\end{figure}

Note that a weighted adjacency graph may admit multiple distinct explanations. An illustrative example is shown on Figure~\ref{fig:manypan}. Two pangenomes may induce exactly the same adjacency multiplicities while differing substantially in their constituent genomes and core marker sets.

\paragraph{Relationship to classical graph problems.} The pangenome peeling problem shares similarities with several well-studied graph decomposition problems. Recall that a path cover of a graph $G$ is a collection of pairwise vertex-disjoint paths whose union contains every vertex of $G$. In the minimum path cover problem one seeks a path cover of minimum cardinality. Variants requiring all paths to be non-trivial have also been studied~\cite{nontrivialpc}. At first glance, constituent genomes resemble the paths appearing in a non-trivial path cover, since every genome is itself a path forest. However, an important distinction is that pangenome peeling operates on weighted adjacencies and allows markers to occur in multiple constituent genomes. Consequently, the genomes explaining a weighted adjacency graph are not required to be vertex-disjoint. The problem is also related to factor problems in graph theory. Given constants $a$ and $b$, an $[a,b]$-factor of a graph is a spanning subgraph in which every vertex has degree between $a$ and $b$. In particular, a $[1,2]$-factor corresponds to a spanning collection of non-trivial paths and cycles. While polynomial-time algorithms are known for finding such structures~\cite{nontrivialpc}, these formulations do not account for edge multiplicities or overlapping path forests. Therefore, existing path-cover and factor techniques cannot be applied directly to the pangenome peeling problem.

\section{Bounds on the Number of Constituent Genomes}

Before attempting to reconstruct constituent genomes, it is useful to determine how many genomes may be required to explain a weighted adjacency graph. This quantity is generally unknown and depends on how adjacency multiplicities can be distributed across path forests. We therefore begin by deriving lower and upper bounds on the size of any pangenome that explains a given weighted adjacency graph. Given an adjacency $(u,v)$, its residual multiplicity is the number of occurrences of $(u,v)$ that remain to be assigned to constituent genomes during reconstruction. The following simple construction provides an immediate upper bound:

\begin{lemma}{(Upper bounds)}
    Let $H = (\mathcal{G}. A. m)$ be a weighted adjacency graph. Then there exists a pangenome $P$ explaining $H$ with at most $\sum_{(u,v) \in A}m(u,v)$ genomes.
\end{lemma}

To see this, create one genome consisting of the single adjacency $(u,v)$ for every occurrence counted by $m(u,v)$. The resulting collection explains every multiplicity exactly.

Obtaining a lower bound, however, is considerably less straightforward. Before considering the setting where multiplicities are different, we will focus on the special case where every adjacency has multiplicity one.

In this case, explaining a weighted adjacency graph $H$ with $k$ genomes is equivalent to decomposing $H$ into $k$ linear forests. Consequently, the minimum number of constituent genomes corresponds exactly to the linear arboricity of $H$, denoted $\mathrm{la}(H)$. A classical argument yields the following lower bound $\mathrm{la}(H) \geq \left\lceil \frac{\Delta(H) + 1}{2} \right\rceil$, where $\Delta(H)$ denotes the maximum degree of $H$~\cite{akiyama1980covering}. The intuition behind this is that each linear forest contributes at most two incident edges to any vertex. This idea will help us understand the general case.

In the general case, for a fixed gene $v\in\mathcal G$, each constituent genome contributes at most two incident adjacencies to $v$. The following lemma formalizes this fact.

\begin{lemma}
    \label{lem:localk}
    Let $H=(\mathcal G,A,m)$ be a weighted adjacency graph and let $P$ be a pangenome of size $k$ explaining $H$. Then for every $v\in\mathcal G$:
    \begin{equation}
        k \geq \left\lceil \frac{\sum_{u\in N_H(v)} m(u,v)} {2} \right\rceil.
    \end{equation}
\end{lemma}

The bound of Lemma~\ref{lem:localk} captures only local degree constraints and therefore does not completely characterize the global structure of $H$. For example, if $H$ is a cycle, every vertex has degree two and the lemma yields a lower bound of one genome. However, a cycle is not itself a path forest and therefore cannot be explained by a single genome. Two genomes are required. Thus global topological constraints may increase the minimum pangenome size beyond the value predicted by local degrees alone.

The previous lemma provides a simple degree-based lower bound. In the weighted setting, however, a vertex may participate in several copies of the same adjacency. We therefore derive a stronger local estimate by considering how adjacency multiplicities can be paired within constituent genomes. We formalize this idea for the general weight case in Algorithm~\ref{alg:bounds}.

\begin{algorithm}
    \DontPrintSemicolon
    \SetKwProg{Fn}{ Function}{}{end}
    \SetKwFunction{FLocal}{LocalK}
    \Fn{ \FLocal{$v,H$}}
    {
    Let $N_H(v) = \{v_1,v_2,\dots,v_j\}$.\;
    Let $R_v = \{r_1,r_2,\dots, r_j\}$, where $r_i = m(v,v_i)$, be the set of multiplicities of $N_H(v)$.\;
    Initialize $k_v \leftarrow 0$.\;
    \While{ $R_v$ contains at least two non-zero entries \label{line:bwhile}}{
    Let $r_1$ and $r_2$ be the two entries with the largest multiplicities in $R_v$.\;
    Set $p = \min(r_1,r_2)$\;
    $k_v \leftarrow k_v + p$\;
    Set $r_1 \leftarrow r_1 - p$ and $r_2 \leftarrow r_2 - p$\;
    }
    \If{one non-zero entry $r$ remains in $R_v$}{
    $k_v \leftarrow k_v + r$
    }
    \Return $k_v$.\;
    }
    \caption{ Greedy pairing computation.}
    \label{alg:bounds}
\end{algorithm}


The value returned by Algorithm~\ref{alg:bounds} corresponds to the minimum number of genomes required to accommodate all incident adjacency occurrences of a single gene under the degree-two constraint.

Once the local bounds are computed for all vertices, we can estimate an approximate lower bound of $H$ as:
\begin{equation}\label{eq:bounds}
        k_v = \max_{v \in \mathcal{G}}\{k_v | k_v = LocalK(v,H) \forall v \in \mathcal{G}\}
\end{equation}

The bound on Equation~\ref{eq:bounds} can be computed efficiently. For a vertex $v$, Algorithm~\ref{alg:bounds} performs $deg_H(v)-1$ pairing operations, since each iteration of Line~\ref{line:bwhile} eliminates at least one non-zero multiplicity. Consequently, evaluating this bound for all vertices takes polynomial time.

In the unweighted case, the bound above reduces to the classical degree lower bound for linear arboricity. In the weighted setting, it naturally extends this idea by accounting for adjacency multiplicities. Although still local in nature and therefore not always tight, it provides an inexpensive estimate that is useful for guiding reconstruction algorithms. The bounds above characterize the feasible range of solution sizes but do not provide a constructive procedure for recovering the constituent genomes themselves. We now turn to the pangenome peeling problem and introduce two reconstruction strategies that seek to generate biologically plausible pangenomes while respecting the observed adjacency multiplicities.

\section{Methods for Pangenome Peeling}

Recall that any weighted adjacency graph $H= (\mathcal{G}, A, m)$ admits a trivial explanation: for every adjacency $(u,v) \in A$. one may create $m(u,v)$ genomes consisting solely of that edge. While such a construction exactly reproduces all observed multiplicities, it yields a highly fragmented pangenome and generally introduces many more genomes than necessary.

Our objective is therefore to join compatible adjacencies into larger path forests whenever possible while preserving the observed multiplicities. Since a weighted adjacency graph may admit multiple valid explanations, we focus on constructing the biologically meaningful solutions rather than attempting to enumerate all possibilities. The methods presented in this section iteratively construct a pangenome while keeping track of the adjacencies that have not yet been fully explained. We therefore introduce two concepts that are used throughout the reconstruction algorithms.

Let $P = \{G_1. G_2, \dots , G_k\}$ be a pangenome and let $(u,v)$ be an adjacency. A genome $G_i$ is a \emph{slack genome} for $(u,v)$ if $(E(G_i) \cup \{(u,v)\}, V(G_i) \cup \{u,v\})$ is also a genome. Equivalently, adding $(u,v)$ does not violate the path forest condition. The set of all slack genomes for an adjacency $(u,v)$ is denoted by $S_{uv}$. For a given edge $(u,v)$, a genome belongs to its slack set if adding $(u,v)$ does not create a vertex of degree greater than two and if it does not create a cycle within a connected component. Thus the slack set identifies all genomes that can still accommodate $(u,v)$ without breaking the path forest constraints.

Let $H=(\mathcal G,A,m)$ be a weighted adjacency graph and let $P$ be a partially reconstructed pangenome. We define the residual multiplicity of an adjacency $(u,v)\in A$ as $r_{uv}=m(u,v)-w_P(u,v)$. We refer to the set of all residual multiplicities as \emph{residuals}. Intuitively, $r_{uv}$ represents the number of occurrences of $(u,v)$ that still need to be assigned to constituent genomes so that $P$ explains $H$.

\subsection{Sequential Edge Addition Algorithm (SEQ)}
Our first approach processes adjacencies in decreasing order of multiplicity. For each occurrence of an adjacency, the algorithm aims to place it into an existing slack genome. If no such genome exists, a new genome containing that adjacency is created. Intuitively, the procedure greedily reuses already existing genomes whenever doing so preserves the path-forest property. New genomes are introduced only when all existing genomes would become invalid after the insertion of the edge. This procedure is formally presented in Algorithm~\ref{alg:seq}.

\begin{algorithm}
    \DontPrintSemicolon
    \SetKwProg{Fn}{ Function}{}{end}
    \SetKwFunction{FSeq}{SequentialEdgeAddition}
    \Fn{ \FSeq{$H = (\mathcal{G},A,m)$}}
    {
    Compute the residuals $r_{u,v}$ for all $(u,v) \in A$.\;
    Let $R = (r_1,r_2,\dots, r_m)$, be a sorted sequence of residuals where $r_i \geq r_{i+1} $.\;\label{line:residuals}
    Initialize $P \leftarrow \emptyset$.\;
    \ForEach{ $r \in R$ \label{line:forres}}{
    Compute the slack set $S_{uv}$ from $P$.\; \label{line:slackset}
    \While{ $r > 0$ \label{line:whileres}}{
    \eIf{ $S_{uv} \neq \emptyset$}{
    Let $i$ be the first element of $S_{uv}$.\;
    $G_i \leftarrow (E(G_i) \cup \{(u,v)\}, V(G_i) \cup \{u,v\})$.\; \label{line:slackassign}
    Remove $i$ from $S_{uv}$.\;
    }{
   Create a new genome $G^*$ such that $E(G^*) = \{(u,v)\}$ and $V(G^*) = \{u,v\}$.\;
   Set $P \leftarrow P \cup \{G^*\}$.\;
    }
    }
    }
    
    \Return $P$.\;
    }
    
    \caption{ Sequential Edge Addition (SEQ) Algorithm.}
    \label{alg:seq}
\end{algorithm}

\begin{lemma}
    Let $H = (\mathcal{G}, A, m)$ be a weighted adjacency graph. The pangenome $P$ given by Algorithm~\ref{alg:seq} explains $H$.
\end{lemma}

\begin{proof}
    We claim that throughout the execution of Algorithm~\ref{alg:seq} the following two conditions are satisfied: every genome in $P$ is a valid path forest and the number of copies of an inserted adjacency $(u,v)$ never exceeds its required multiplicity.
    
    Initially $P = \emptyset$ so both properties hold trivially.
    Consider an adjacency $(u,v) \in A$ with multiplicity $m(u,v)$.
     The outer loop in Line~\ref{line:forres} processes every adjacency of $H$ while the inner loop in Line~\ref{line:whileres} executes exactly $m(u,v)$ times. During each iteration, one occurrence of $(u,v)$ is inserted into a genome. If a slack genome exists, $(u,v)$ is inserted into that genome. By definition of slack genome, the resulting graph is a valid genome. Otherwise, a new genome consisting only of the adjacency $(u,v)$ is created. Therefore every insertion preserves the invariant that all the constituent genomes of $P$ are valid genomes. 
    
    Note that the residuals computed in Line~\ref{line:residuals} correspond to given multiplicities. After all residuals have been processed, exactly $m(u,v)$ copies of every adjacency $(u,v) \in A$ have been inserted across the genomes of $P$. Therefore, the final endogenous weight $w(u,v)$ of every adjacency satisfies $w(u,v) = m(u,v)$. Hence, the resulting pangenome explains $H$.
\end{proof}

\begin{lemma}
    \label{lem:seq_time}
    Let $H = (\mathcal{G}, A, m)$ be a given weighted adjacency graph such that $|\mathcal{G}| = n$. Let $M = \sum_{(u,v) \in A} m(u,v)$. Algorithm~\ref{alg:seq} can be implemented in $O(M^2 n)$ time.
\end{lemma}

\begin{proof}
    The while loop on Line~\ref{line:whileres} inserts exactly one adjacency into a genome. Since the total number of adjacency occurrences is $M$, the algorithm performs exactly $M$ insertions. For each insertion, the most costly operation is the slack set computation on Line~\ref{line:slackassign}. One can compute the slack set by scanning all genomes and testing whether $(u,v)$ can be inserted into each genome. Degree verifications can be done in constant time using adjacency information, while cycle detection requires determining whether $u$ and $v$ belong already to the same path component. Note that every genome $G_i$ satisfies $|V(G_i)| \leq n$. Thus the slack set verifications can be done in $O(n)$ for a genome $G_i$. Note that at most one genome can be created per insertion, we have then that $k \leq M$, thus each insertion requires $O(Mn)$ time. Thus over all the $M$ insertions, the total running time is $O(M^2n)$. Hence, Algorithm~\ref{alg:seq} runs in polynomial time.
\end{proof}

\begin{figure}
    \centering
    \includegraphics[width=\textwidth]{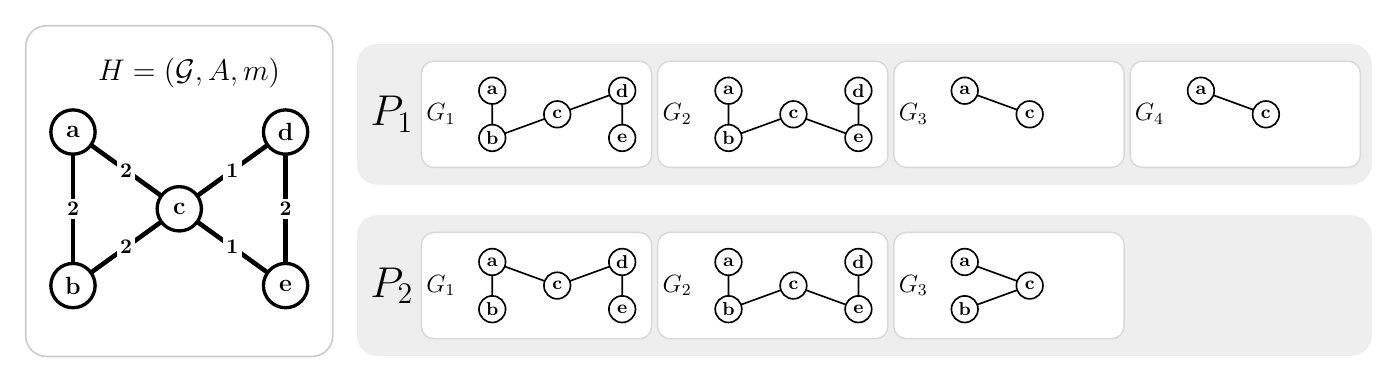}
    \caption{ Algorithm~\ref{alg:seq} might add more genomes than necessary. Using the given instance $H$, the pangenome $P_1$ is one possible output of the sequential addition algorithm that proceeded by adding the edges in the following order $(a,b),(b,c),(a,c),(d,e),(c,d),(c,e)$. $P_2$ is an optimal solution that takes one less genome.}
    \label{fig:counter_naive}
\end{figure}

Although the algorithm yields a pangenome that explains $H$, this pangenome is not necessarily a pangenome of minimum size. A counter example of this is presented in Figure~\ref{fig:counter_naive}.

It should be pointed out that Algorithm~\ref{alg:seq} guarantees a feasible solution but does not guarantee optimality with respect to the number of constituent genomes. Greedy insertion decisions made early in the construction may prevent subsequent adjacencies from being incorporated into existing genomes, forcing the creation of additional genomes. Figure~\ref{fig:counter_naive} illustrates such a situation. The pangenome $P_1$ corresponds to a feasible output of the sequential edge addition algorithm, whereas $P_2$ explains the same weighted adjacency graph using one fewer genome.

\subsection{ Maximum Spanning Tree Peeling Algorithm (MST) }

The SEQ algorithm treats each adjacency independently and therefore makes decisions solely based on local properties. These may lead to fragmented solutions and extravagant creation of additional genomes.

To obtain a more structured reconstruction, we exploit the observation that a tree is already very close to a path forest. In particular, if highly supported adjacencies are selected first, they can serve as a backbone from which longer genomic paths can be reconstructed. This motivates the use of a maximum spanning tree (MST) of the weighted adjacency graph. Given a weighted adjacency graph we first construct a maximum spanning tree $T$. Since $T$ retains the adjacencies of highest multiplicity, it captures the strongest structural signal present in the input graph. Paths extracted from $T$ are therefore expected to correspond to frequently observed marker orderings.

Our proposed algorithm consists of three phases:
\begin{enumerate}
    \item \textbf{MST Peeling.} A maximum spanning tree is constructed for each connected component of the weighted adjacency graph. The tree is then traversed bottom-up and paths are progressively extended along high-weight adjacencies.
    \item \textbf{Base Pangenome Construction.} The paths extracted from the spanning tree are merged into genomes while preserving the path-forest constraint. This phase forms the base pangenome.
    \item \textbf{Weight reconciliation.} Note that the spanning tree contains only a subset of the adjacencies present in $H$. The residual multiplicities are therefore reinserted using the same slack-genome strategy described in the sequential edge addition algorithm. This guarantees that the final endogenous edge weights match the observed multiplicities.
\end{enumerate}

\begin{figure}
    \centering
    \includegraphics[width=\textwidth]{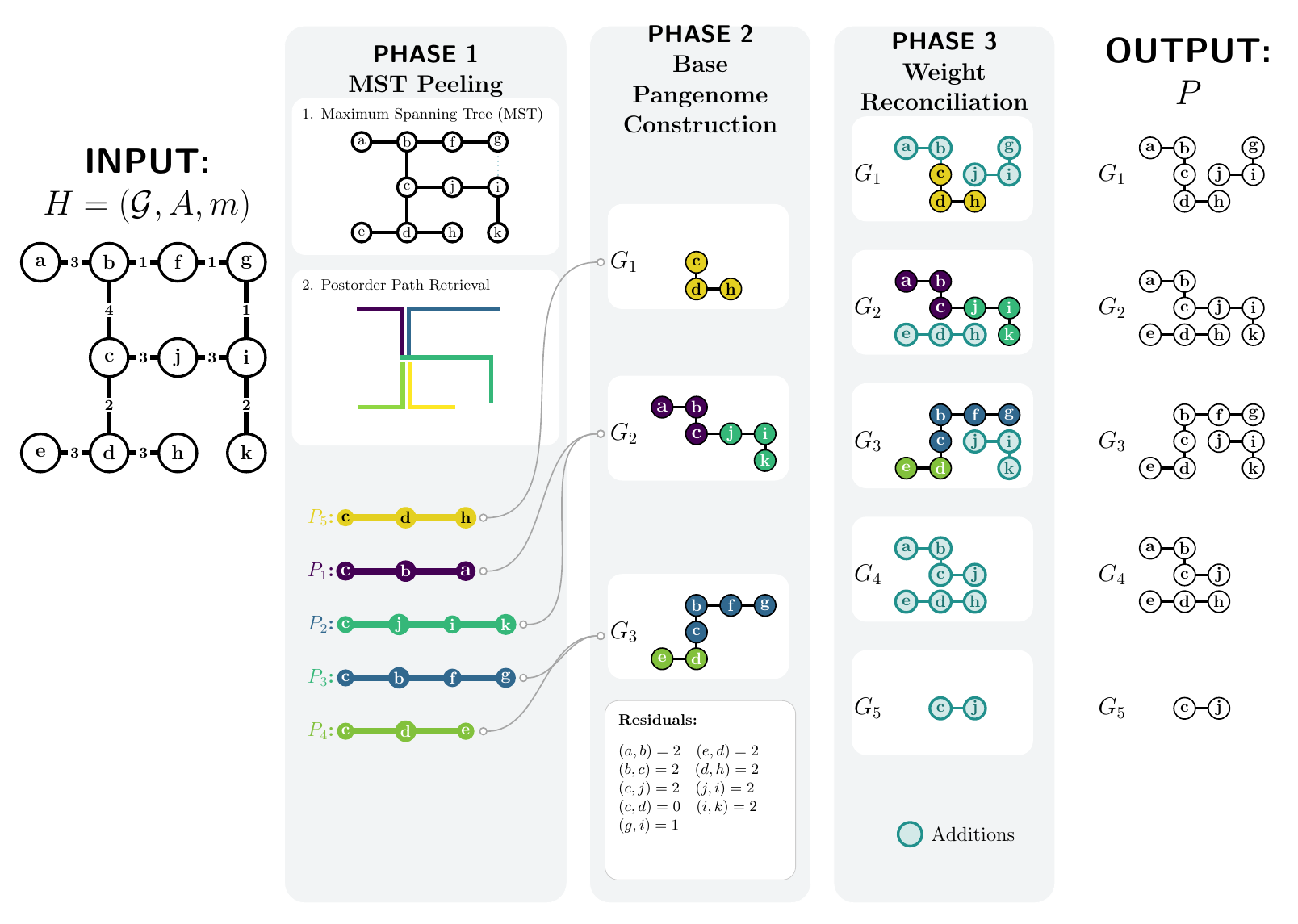}
    \caption{An illustrative example of a solution using Algorithm~\ref{alg:mst} and input graph of Figure~\ref{fig:manypan}. In the first phase, we compute the Maximum Spanning Tree, then do a post order path retrieval per leaf. On the second phase, we merge the obtained paths to construct a Base Pangenome. On phase 3, we add all missing edges whose residuals are more than cero, creating new genomes when current ones do not admit a given adjacency.}
    \label{fig:mstpipeline}
\end{figure}

A complete worked example of the pipeline is shown in Figure~\ref{fig:mstpipeline}.

\paragraph{Phase 1: MST Peeling.} Let $T$ be a maximum spanning tree of a connected component of $H$. Starting from the leaves of $T$, we perform a postorder traversal. Each leaf initially contributes a path consisting of a single marker. When processing an internal node $v$, every path produced by a child $u$ is extended through the adjacency $(u,v)$ whenever a positive residual multiplicity remains for that edge. Each successful extension reduces the residual weight by one. Consequently, high multiplicity edges may contribute to multiple reconstructed paths. At the end of the traversal, the last visited vertex contains a collection of paths.

\paragraph{Phase 2: Base Pangenome Construction.} The paths produced during MST peeling must subsequently be converted into genomes. Two paths can be assigned at the same genome whenever they do not violate the path-forest constraints. In particular paths might be joined if their marker sets are disjoint or if they intersect in a single marker whose degree remains at most two after insertion. The resulting base pangenome forms an initial explanation of the spanning-tree backbone. Since the MST contains only the most supported adjacencies, these genomes tend to contain substantially longer paths than those produced by the purely sequential strategy.

\paragraph{Phase 3: Weight reconciliation.} The MST reconstruction explains only the  multiplicities represented by the spanning tree. For every adjacency $(u,v)$ with positive residual multiplicity, we repeatedly insert additional occurrences of $(u,v)$ into slack genomes whenever possible. If no slack genome exists, then a new genome is created. This reconciliation continues until all residual multiplicities vanish. Since every residual occurrence is ultimately assigned to some genome, the final pangenome explains the original weighted adjacency graph exactly.

\begin{algorithm} 
\DontPrintSemicolon 
\SetKwProg{Fn}{Function}{}{end} 
\SetKwFunction{FPeel}{ MSTPeel} 
\SetKwFunction{FMerge}{ MergePaths} 
\SetKwFunction{FMST}{ MSTPeeling} 
\Fn{ \FPeel{$T,R$} \label{line:peelphase}}{ 
Let $Q_v$ denote the set of paths associated to $v$.\;
\ForEach{$v \in T.postorder()$}{ 
    \eIf{$v$ is a leaf
    }{
    Create path $p$ that consists of the vertex $v$.\;
    Add $p$ to $Q_v$
    }{ 
    Initialize $Q_v \leftarrow \emptyset$.\;
    \ForEach{
    $u \in children(v)$
    }{ 
        \ForEach{$p \in Q_u$ where $u$ belongs to an endpoint of $p$}{ 
            \If{$r_{uv} > 0$
            }{ 
            Extend $p$ by appending $v$ and update the endpoint.\;
            $r_{uv} \leftarrow r_{uv}-1$.\; 
            } 
            Add $p$ to $Q_v$.\; 
            } 
    } 
    } 
    } 
    \tcp{Note that $v$ is the last visited vertex}
    \Return $Q_v$.\; 
    } 
    \vspace{1ex} 
    \Fn{ \FMerge{$\mathcal{Q}$ \label{line:MergePaths}}}{ 
    Initialize $P \leftarrow \emptyset$.\; 
    \ForEach{path $q \in \mathcal{Q}$ \label{line:qslack}}{ 
        \If{there exists a  slack genome $G \in P$ for an endpoint of $q$.
        }{ 
            Insert $q$ into $G$.\; 
        } 
        \Else{ 
            Create a new genome containing $q$.\; 
            Add it to $P$.\; 
            } 
        } 
    \Return $P$.\; 
    } 
    \vspace{1ex} 
    \Fn{\FMST{$H=(\mathcal{G},A,m)$}}{ 
    Compute the connected components of $H$.\; 
    Let $\mathcal P \leftarrow \emptyset$.\; 
    \ForEach{connected component $C$}{ 
    Compute a maximum spanning tree $T$ of $C$.\; 
    Initialize residuals $r_{uv}\leftarrow m(u,v)$ for every adjacency of $C$.\; 
    $\mathcal Q \leftarrow$ \FPeel{$T,R$}.\; \label{line:MSTpaths}
    \tcp{Build genomes for component $C$} $P_C \leftarrow$ \FMerge{$\mathcal Q$}.\; 
    Add $P_C$ to $\mathcal P$.\; } 
    \tcp{Merge genomes from distinct connected components} 
    Select one set $P^\star \in \mathcal P$.\; 
    \ForEach{$P_C\in \mathcal P\setminus\{P^\star\}$}{ 
    Merge genomes of $P_C$ into genomes of $P^\star$ whenever possible.\; \label{line:compmerge}
    } 
    $P \leftarrow P^\star$.\; 
    \tcp{Weight reconciliation } 
    \ForEach{$(u,v)\in A$ such that $r_{uv}>0$ \label{line:weightline}}{ 
    \While{$r_{uv}>0$}{ Insert $(u,v)$ into a slack genome if one exists, otherwise create a new genome.\; 
    $r_{uv}\leftarrow r_{uv}-1$.\; } } 
    \Return $P$.\; }
\caption{Maximum Spanning Tree Peeling Algorithm (MST).} \label{alg:mst} \end{algorithm}

\begin{lemma}\label{lem:mst_correct}
Let $H=(\mathcal{G},A,m)$ be a weighted adjacency graph. Algorithm~\ref{alg:mst} returns a pangenome $P$ that explains $H$. 
\end{lemma}
\begin{proof}
    We claim that throughout the execution of Algorithm~\ref{alg:mst} the following two conditions are satisfied: every genome in $P$ is a valid path forest and no inserted adjacency $(u,v)$ exceeds its required multiplicity.
    
    During the MST peeling phase on Line~\ref{line:peelphase}, paths are extended only through adjacencies with positive residual multiplicity. Whenever an adjacency is used to extend a path, the corresponding residual $r_{u,v}$ is decreased by one. Since vertices are appended only at path endpoints, each constructed structure remains a simple path. In particular, no paths that consist of isolated vertices can be created since $m(u,v) > 0$ for all $(u,v) \in A$ by definition of $m$. Consequently, the collection of paths $Q_v$ on Line~\ref{line:MSTpaths} consists exclusively of path components. When the function on Line~\ref{line:MergePaths} is executed, paths are inserted into an existing genome only if the resulting graph remains a path forest. Otherwise a new genome is created. Note that since these genomes belong to different connected components, their vertex sets are disjoint, and thus the genome merge on Line~\ref{line:compmerge}. Therefore, every genome constructed by the algorithm is a valid path forest, proving our first claim.
    
    To prove the second claim, after the first two phases, some adjacencies may still have positive residual multiplicity. Let $r_{uv} = m(u,v)-w_P(u,v)$ denote the residual multiplicity of $(u,v)$. The reconciliation phase processes every residual occurrence exactly once. Whenever possible, an occurrence of $(u,v)$ is inserted into a slack genome. Otherwise a new genome containing $(u,v)$ is created. By definition of a slack genome, every insertion preserves the path-forest property. The reconciliation phase terminates only when all residuals vanish. Hence, for every adjacency $(u,v)\in A$ we have $w_P(u,v)=m(u,v)$ . Therefore the resulting pangenome explains $H$.
\end{proof}

\begin{lemma}
    \label{lem:mst_time}
    Let $H = (\mathcal{G}, A, m)$ be a given weighted adjacency graph such that $|\mathcal{G}| = n$. Let $M = \sum_{(u,v) \in A} m(u,v)$. Algorithm~\ref{alg:mst} can be implemented in $O(M^2n)$ time.
\end{lemma}

\begin{proof}
The peeling phase in Line~\ref{line:peelphase} performs at most $M$ path extensions. The reason is that each residual occurrence is processed at most once. Every successful extension of a path decreases some residual $r_{uv}$ by one, and the total number of such decrements over the entire algorithm is at most $M$. Consequently, the peeling phase requires $O(M)$ time.

Let $\mathcal Q$ denote the paths returned by the function on Line~\ref{line:MergePaths}. Note that since every leaf on the given connected component contributes with one path, $|\mathcal Q| \leq n$. Each leaf of the MST initializes a single path and no new paths are created thereafter. Since a tree on $n$ vertices has at most $n$
leaves, at most $n$ paths are maintained throughout the procedure.
Furthermore, each path is extended only along tree edges and therefore contains at most $n$ vertices.  As in Lemma~\ref{lem:seq_time}, we know that the slack genome computations can be performed in $O(n)$ time. As a consequence, the for loop in Line~\ref{line:qslack} can be performed in $O(n^2)$ time.

During the component merge phase (Line~\ref{line:MergePaths}), every path $q\in\mathcal Q$ is inspected once. To determine whether $q$ can be inserted into an existing genome, the algorithm may scan all currently constructed genomes. Since at most one new genome can be created for each path, there are at most $M$ genomes. Using the same slack-set computation as in Algorithm~\ref{alg:seq}, testing a genome requires $O(n)$ time. Hence the merge phase requires at most $O(M^2n)$ time.

Finally, during the weight reconciliation phase on Line~\ref{line:weightline}, every remaining residual occurrence is inserted exactly once. Since there are at most $M$ residual occurrences and each insertion may require scanning at most $M$ genomes of size at most $n$, this phase also runs in $O(M^2n)$ time. Combining the bounds for all phases yields an overall running time of $O(M^2n)$. Therefore Algorithm~\ref{alg:mst} runs in  $O(M^2n)$ time.
\end{proof}

Unlike SEQ, which constructs genomes from individual adjacencies, MST first recovers a globally consistent backbone and only subsequently reconciles missing multiplicities, producing pangenomes with characteristically different structural properties.

\section{Results} 

We have seen that there are various ways of constructing a pangenome from a weighted adjacency graph. In practice, we might wish to define objective functions on these constructions. One way to do this is to simulate pangenomes under realistic conditions and to see how well our algorithms performed on their induced adjacency graphs. Some criteria for this evaluation include how closely the number of constituent genomes output by SEQ and MST approximate the number produced by the simulated input, how much of the core genome is recovered and, most important, how closely the output set of constituent genomes resembles the input. Here we describe a simulation experiment and the results of these three evaluations when we apply SEQ and MST.

\subsection{Simulation benchmark} 
To evaluate pangenome peeling under controlled conditions, we generated synthetic pangenomes for which the constituent genomes were known exactly. Each simulated dataset begins with an ancestral genome consisting of a single linear path containing $n$ unique markers. From this ancestor, we generate $k-1$ descendant genomes independently. Each descendant is obtained by copying the ancestral genome and applying a collection of evolutionary events. The number of occurrences of each event type is sampled independently from a Poisson distribution $N_e \sim \mathrm{Poisson}(r_e \cdot b)$, where $r_e$ is the user-specified event rate and $b$ is a branch-length scaling factor. To generate datasets of varying evolutionary level, the branch-length scale is sampled from a noise-dependent interval. Low evolutionary level datasets use scaling factors between 0 and 0.5, medium level datasets use values between 0.5 and 2.0, and high level datasets use values between 3.0 and 5.0. The simulator implements a variety of structural and gene-content modifications, including gene insertions, gene deletions, adjacency deletions, inversions, rearrangements, chromosome fissions, chromosome fusions, and translocations. These operations modify either the gene content or the adjacency structure of the descendant genomes while preserving biologically meaningful linear genome architectures. The average number of simulated event types per evolutionary level can be found in Table~\ref{tab:events}. The resulting collection of genomes constitutes the ground-truth pangenome. To generate an input instance for pangenome peeling, we aggregate all adjacencies observed across the simulated genomes into a weighted adjacency graph, where edge weights correspond to adjacency frequencies.

The simulated event set was designed to capture both gene-content evolution and large-scale genome rearrangements. Gene insertions and deletions alter the presence or absence of markers, whereas inversions, translocations, fissions, fusions, and adjacency modifications alter genome structure without necessarily changing gene content. This combination produces benchmark instances with varying levels of topological and compositional configurations.

\begin{table}[htbp]
\centering
\caption{Average number of simulation events.}
\begin{tabular}{@{}lrrr@{}}
\toprule
Event Type & Low & Medium & High \\ \midrule
Insertions & 11 & 60.2 & 192.5 \\
Edge deletions & 9.7 & 58 & 172.26 \\
Gene deletions & 9.70 & 56.10 & 158.46 \\
Inversions & 8.26 & 55.26 & 180.76 \\
Fissions & 10.50 & 48.90 & 168.93 \\
Fusions & 10.80 & 58.83 & 197.46 \\
Translocations & 22.03 & 102.73 & 360.56 \\ \bottomrule
\end{tabular}
\label{tab:events}
\end{table}

\subsection{On inferring the number of constituent genomes} 
\begin{figure}[t]
    \centering
    \begin{tabular}{ccc}
    \includegraphics[width=0.32\textwidth]{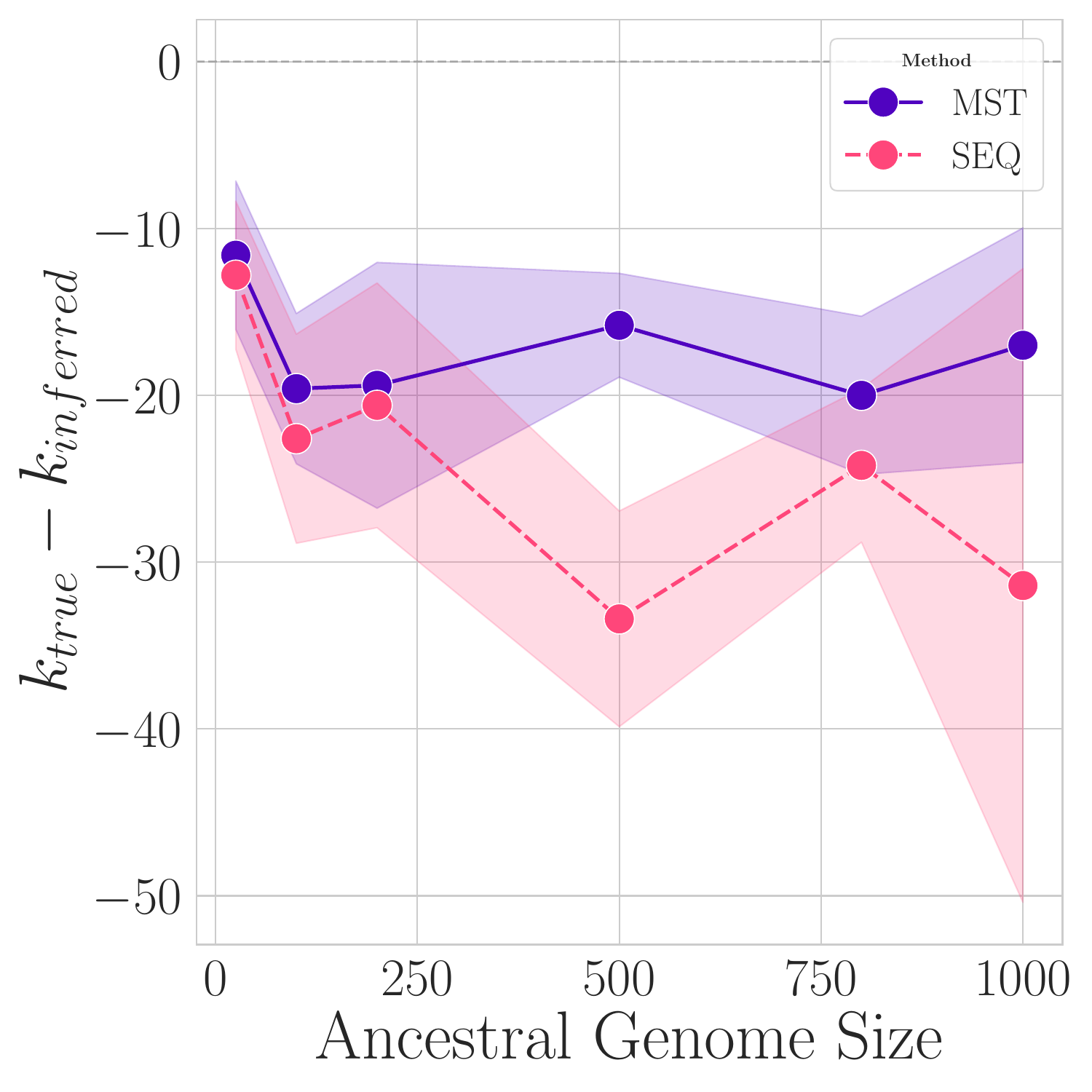}  &
        \includegraphics[width=0.3\textwidth]{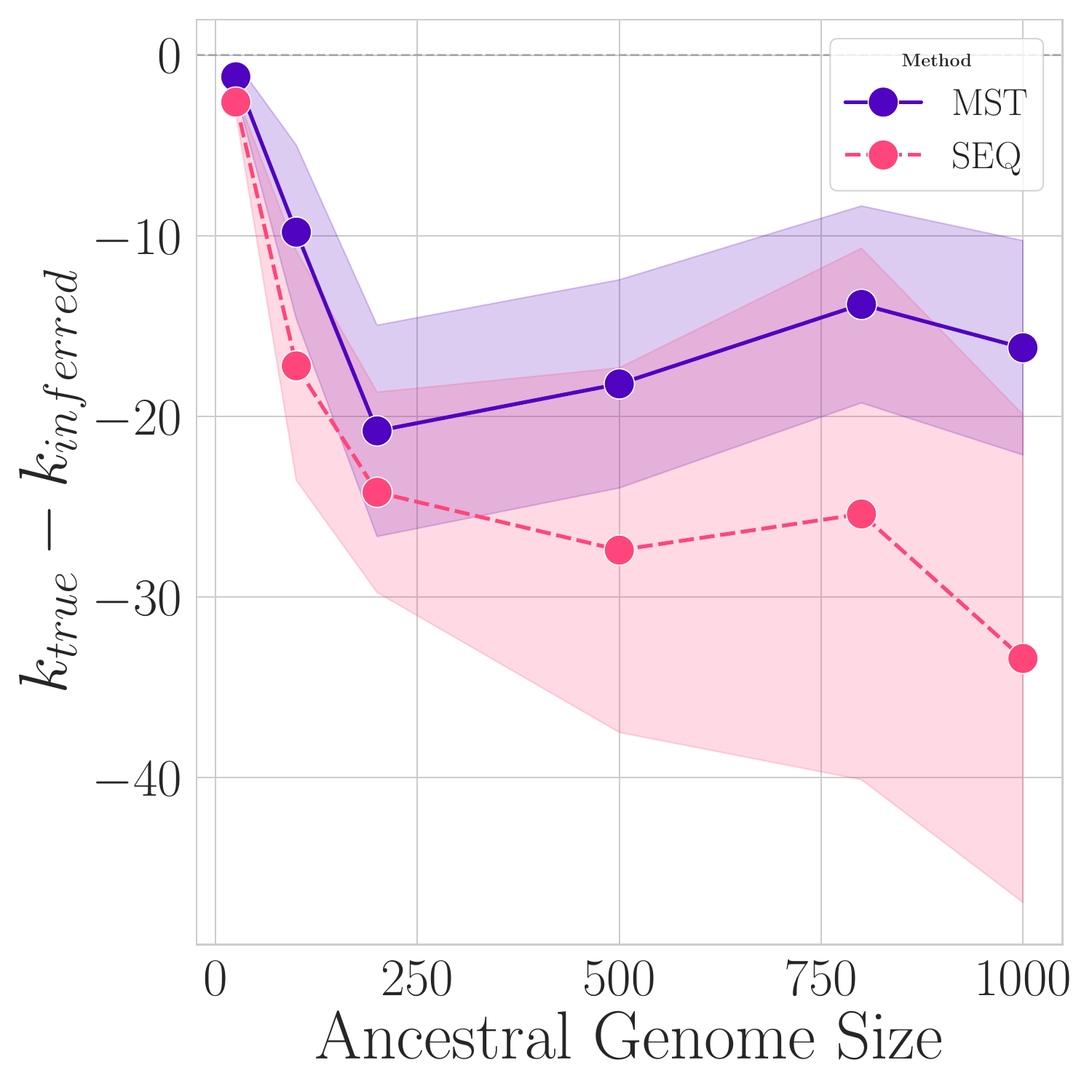}  &
        \includegraphics[width=0.3\textwidth]{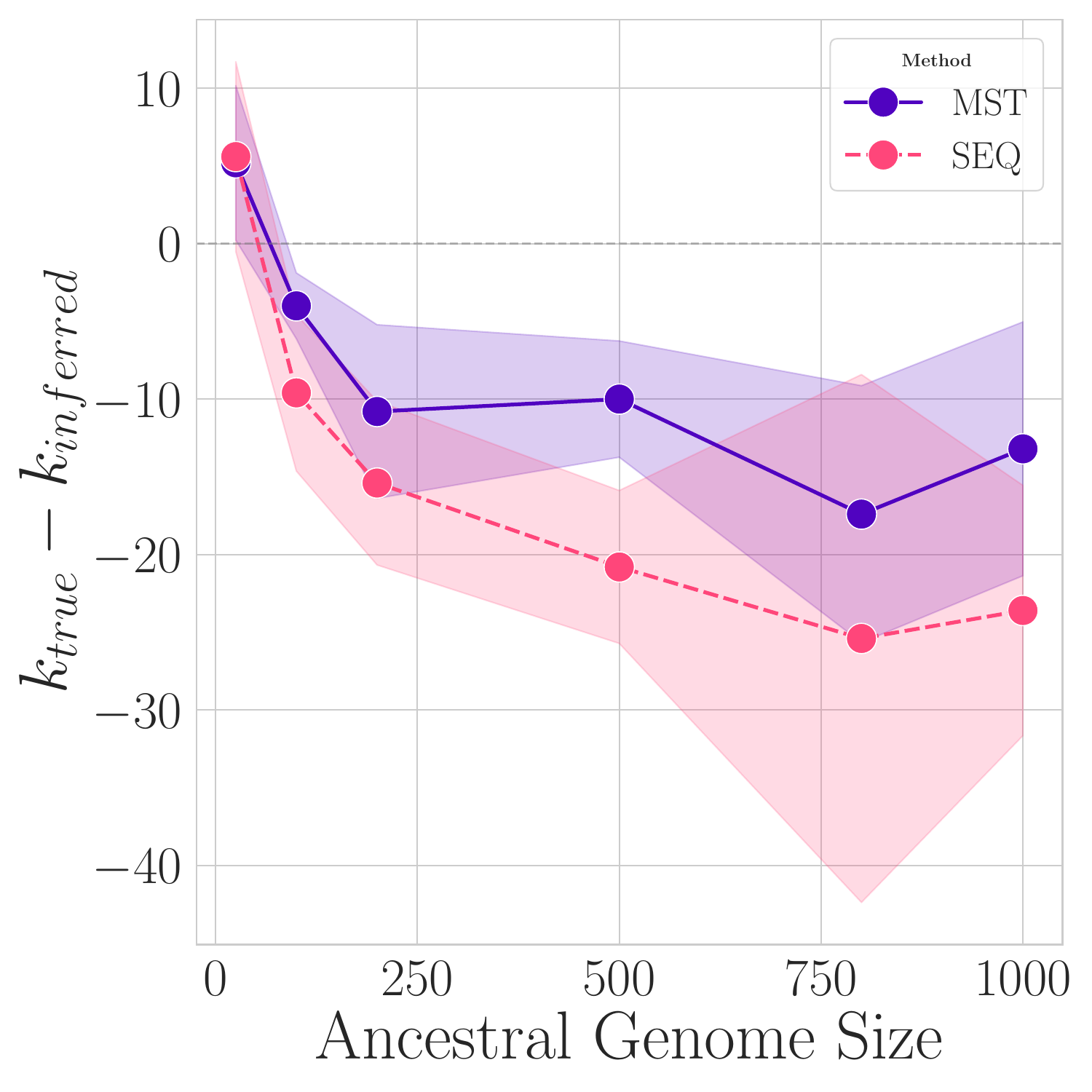}\\
    \includegraphics[width=0.3\textwidth]{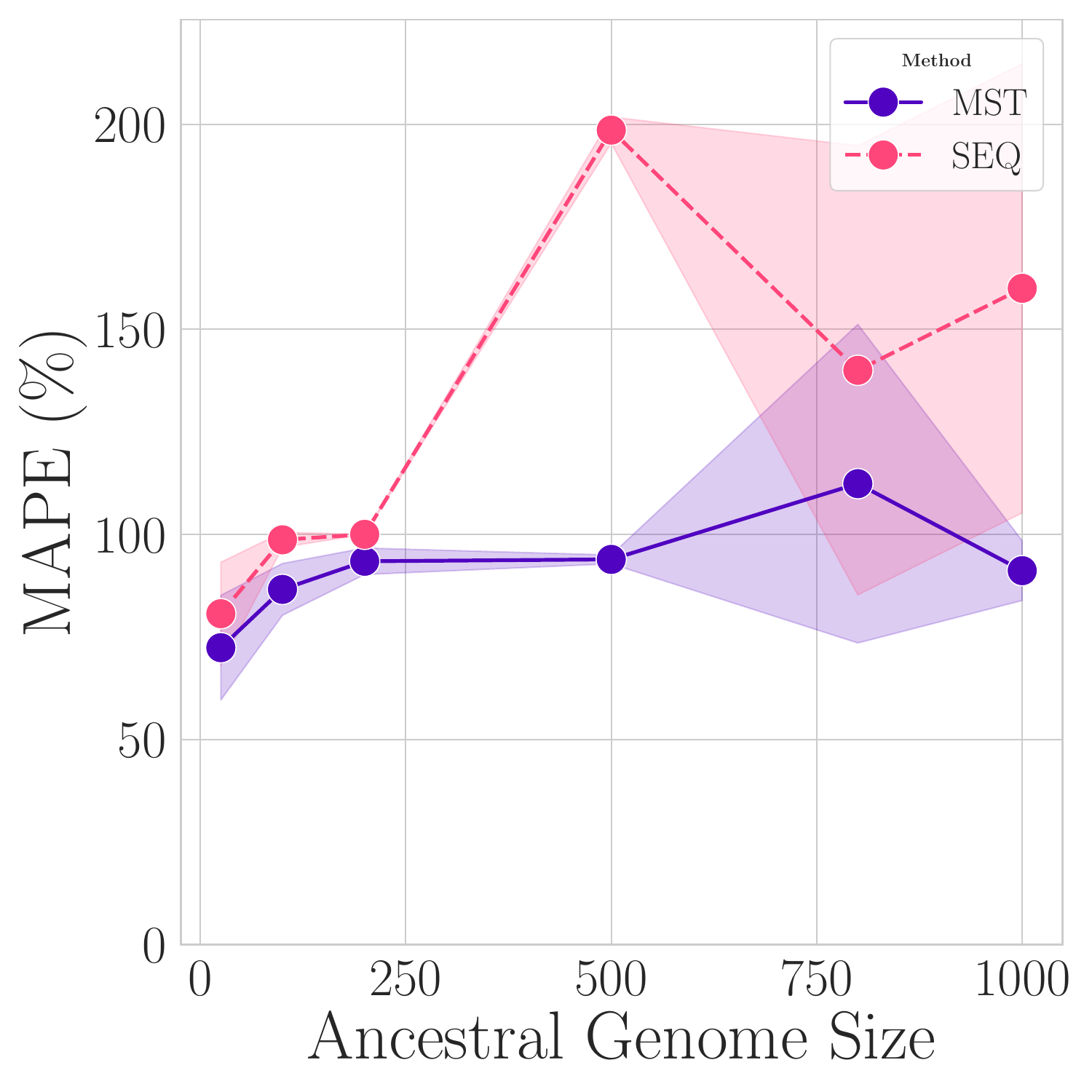}  &
        \includegraphics[width=0.3\textwidth]{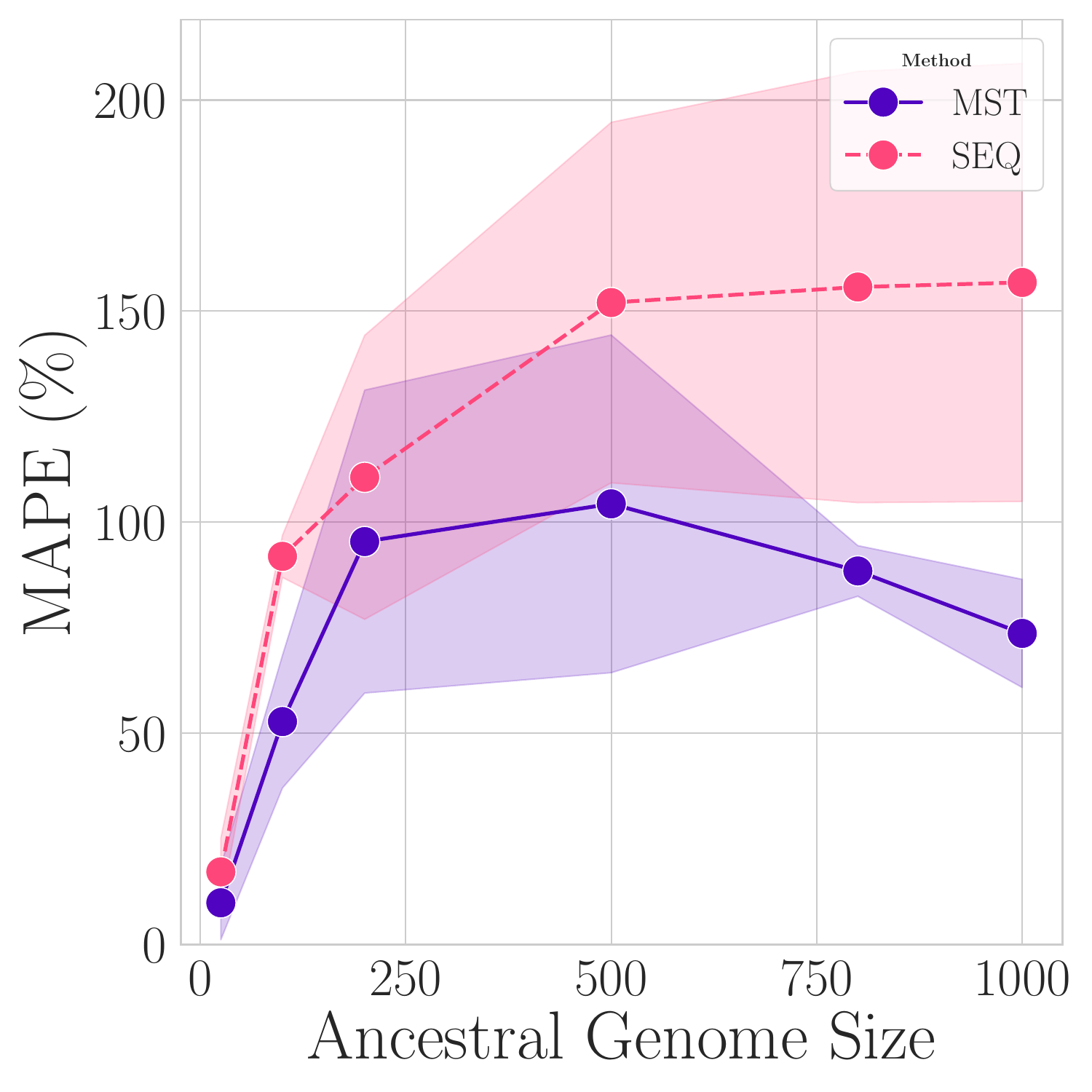}  &
        \includegraphics[width=0.3\textwidth]{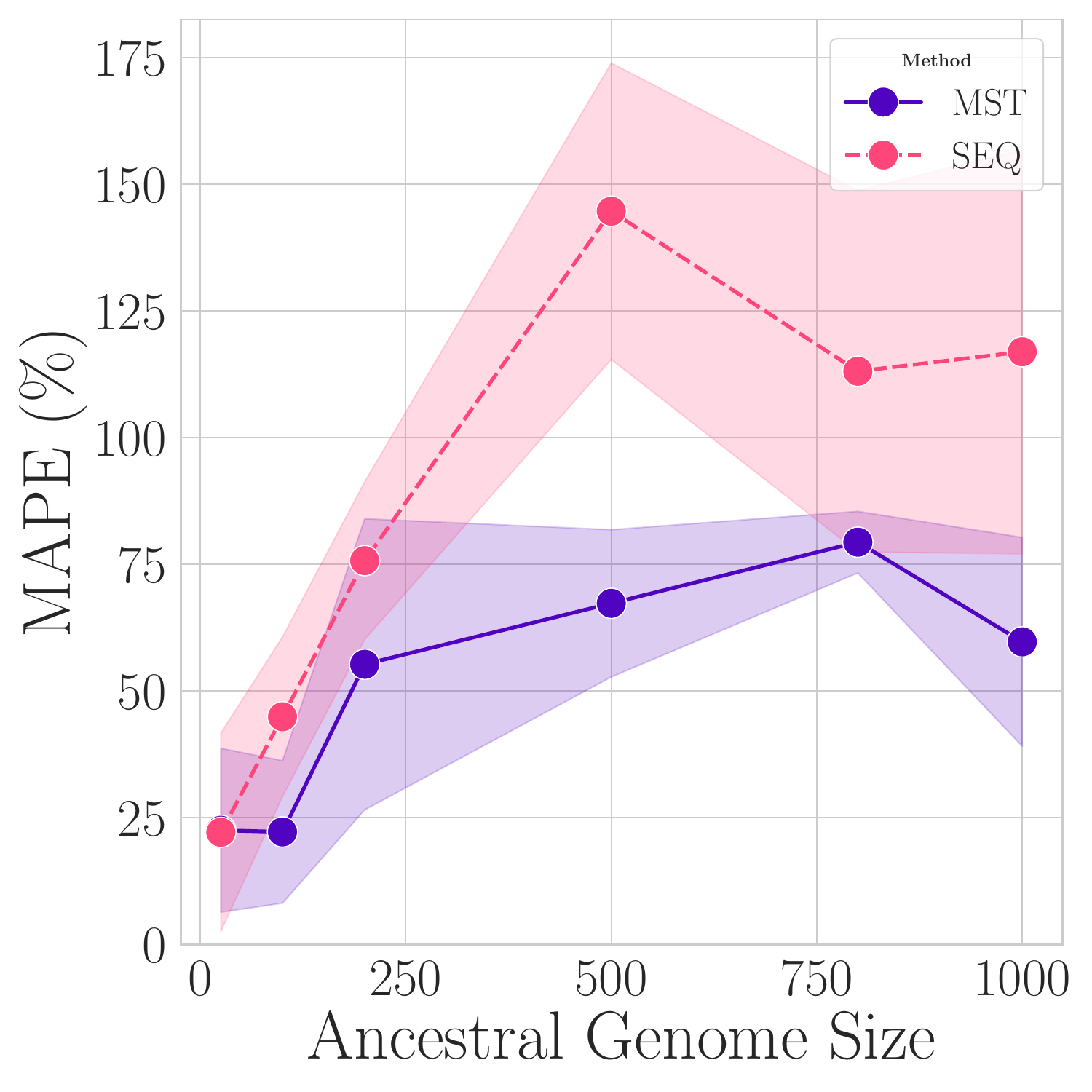}\\
        
        (a) & (b) & (c)  \\
    \end{tabular}
    \caption{  Differences in number of constituent genomes across simulated pangenomes of varying size and complexity. Columns correspond to (a) low , (b) medium and (c) high evolutionary level settings. Figures show the raw estimation error $k_{true} - k_{inferred}$ and their corresponding MAPE values for SEQ and MST methods. On average,  $k_{true} = 19$.} 
    \label{fig:genome_number}
\end{figure}

The first objective of pangenome peeling is to recover the number of genomes that gave rise to the observed weighted adjacency graph. For each simulated dataset, we compared the true number of constituent genomes ($k_{true}$) with the number inferred by each algorithm ($k_{inferred}$). Figure~\ref{fig:genome_number} summarizes the difference between the inferred and the true number of genomes. The top row reports the raw error $k_{true}-k_{inferred}$ while the bottom row is an adapted Mean Absolute Percentage Error (MAPE) calculated as $\mathrm{MAPE} = \frac{|k_{true}-k_{inferred}|} {k_{true}} \times 100$. Overall, MST consistently outperforms SEQ. Across all the different generated instances MAPE values for MST remain approximately below 150\%, while SEQ frequently exceeds this threshold. The best performance of MST occurs in the high evolutionary level simulations, where the average MAPE value remains below 75\%.

\subsection{Inference of core content} 
\begin{figure}[t]
    \centering
    \begin{tabular}{ccc}
        \includegraphics[width=0.3\textwidth]{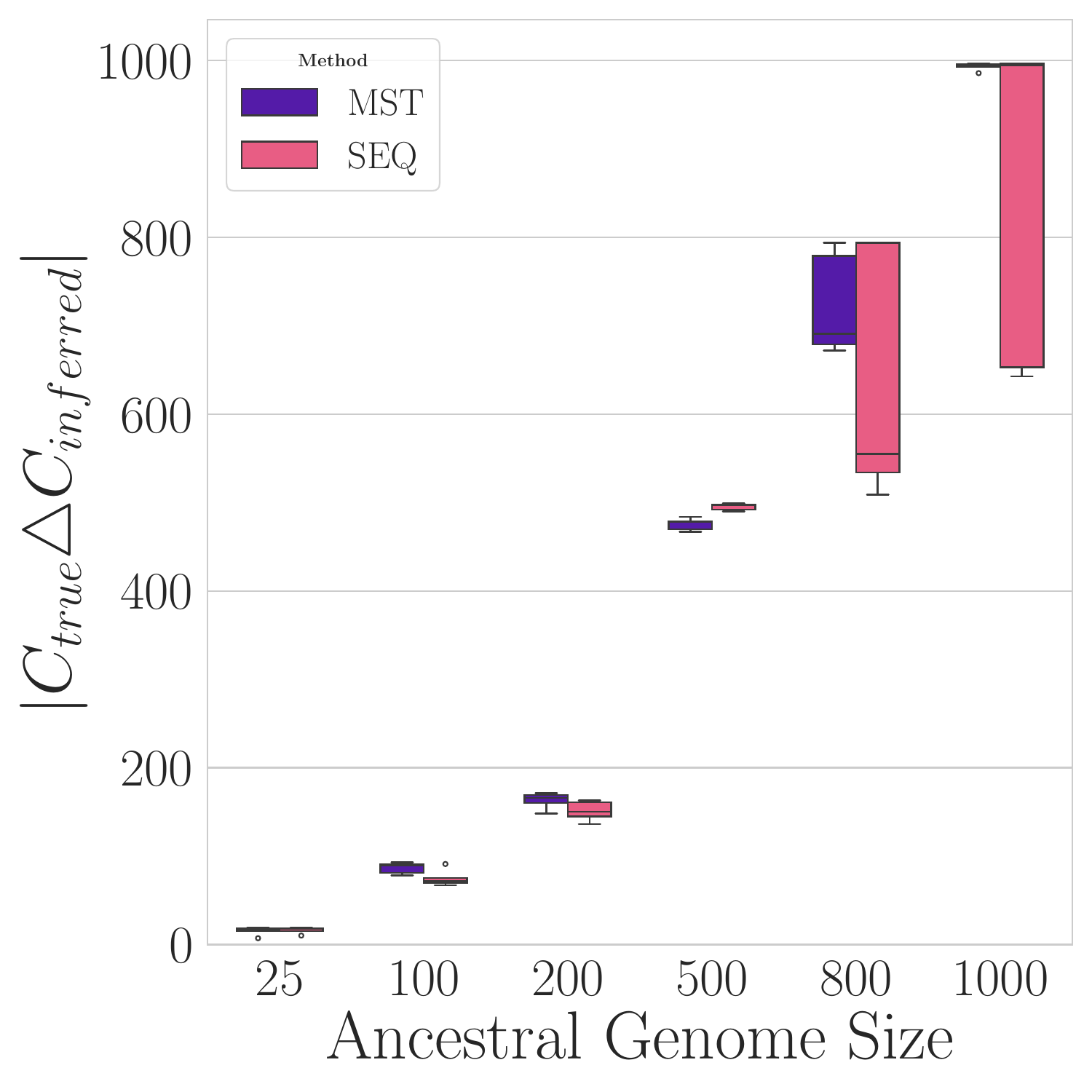}  &
        \includegraphics[width=0.3\textwidth]{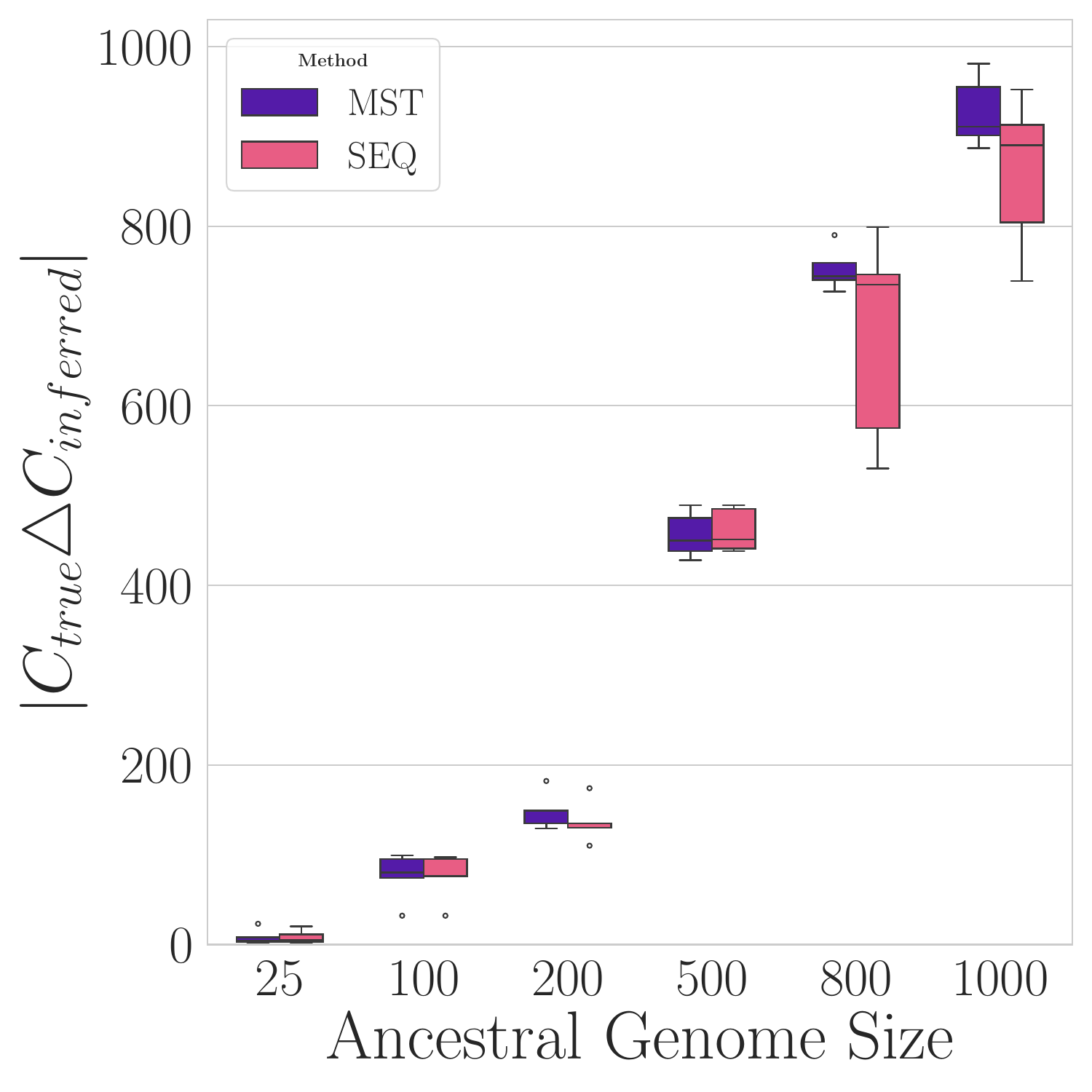}  &
        \includegraphics[width=0.3\textwidth]{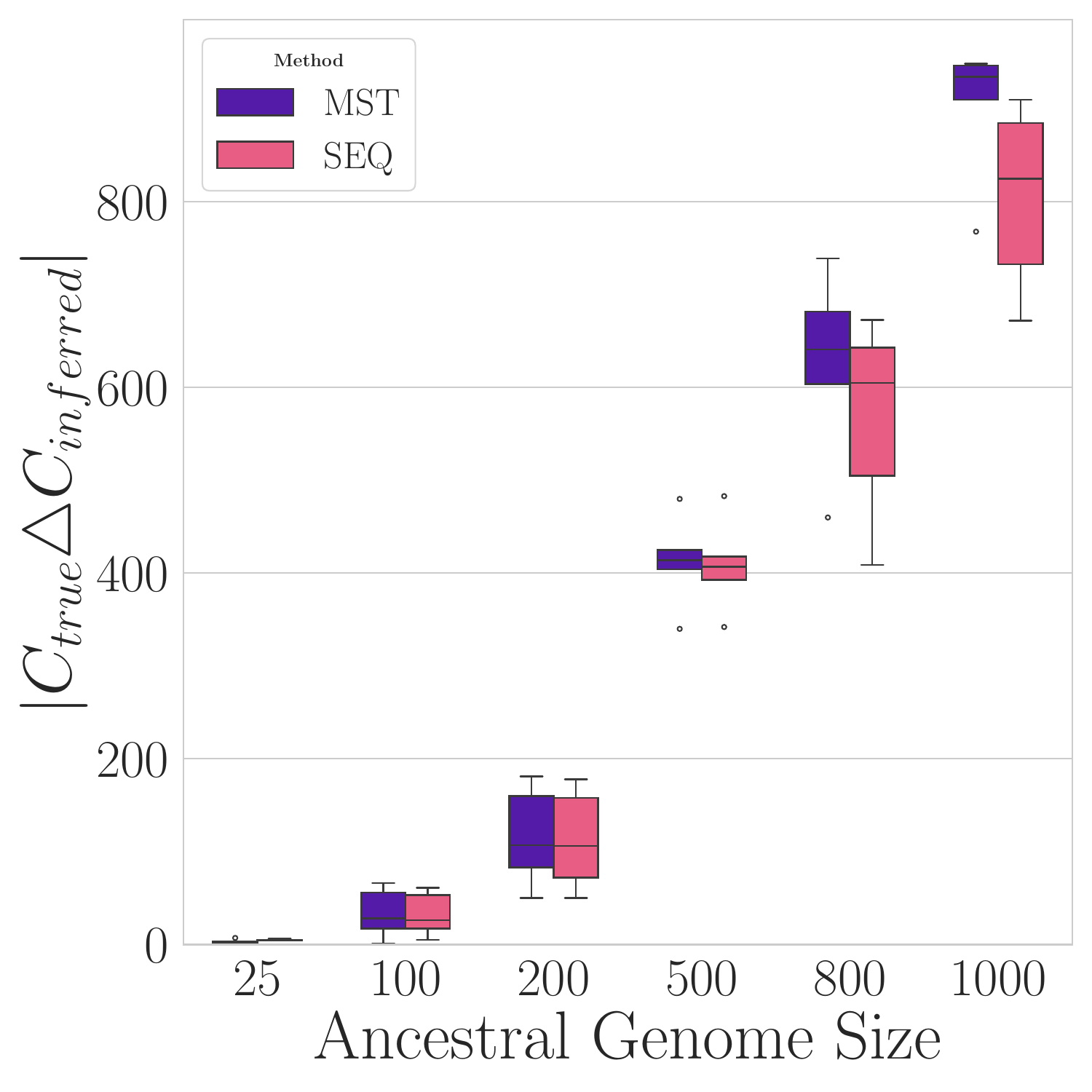}\\
        (a) & (b) & (c)\\
    \end{tabular}
    \caption{  Size of the symmetric difference between the true core gene set and the core gene set of the inferred pangenomes. Columns correspond to different evolutionary levels: (a) low, (b) medium and (c) high. Lower values indicate an accurate recovery of the true core gene set. Results are shown for both methods Sequential Edge Addition (SEQ) and Maximum Spanning Tree (MST). On average, the value of $|C_{true}| = 9$.} 
    \label{fig:core_difference}
\end{figure}

In addition to recovering the number of constituent genomes, we evaluated whether the inferred pangenomes preserved the core set of the simulated datasets. For each benchmark instance, we identified the core as the set of markers present in every constituent genome. We then compared the core of the simulated pangenome with the core set of each inferred solution. Similarity was measured using the symmetric difference between the two sets, $|C_{true} \triangle C_{inferred}|$, where $C_{true}$ denotes the core gene set of the simulated pangenome and $C_{inferred}$ denotes the core gene set reconstructed by each peeling algorithm. A symmetric difference of zero indicates perfect recovery of the core of the simulated instance, whereas larger values indicate markers that were either incorrectly included in or excluded from the inferred core set. Figure~\ref{fig:core_difference} summarizes the results across all simulation settings. 

We observe that for both methods the symmetric difference increases with input size, suggesting that recovering the correct core genome becomes progressively more difficult as the number of markers and constituent genomes grows. However, the two methods exhibit different distributions. As the input size increases, the boxplots diverge not only in their median values but also in their variability and spread, suggesting that both methods differ in both accuracy  and roboustness with respect to core set reconstruction. These differences become more evident in bigger instances.

\subsection{Inference of genome structure} 
Beyond core content and constituent genomes, we evaluated whether the inferred genomes recovered the original simulated set.  Pairs of individual genomes like the ones we simulated are most appropriately compared using the DCJ-indel distance implemented in \texttt{UniMoG}~\cite{unimog}. Unlike the classical DCJ distance, the DCJ-indel formulation accommodates genomes with unequal gene content while remaining symmetric and computationally efficient~\cite{indels}. The distance accounts for insertions, deletions, inversions, translocations, fusions, and fissions.

The comparison between the two sets of genomes, one simulated and the other inferred by peeling of the adjacency graph, is complicated by the lack of any necessary one-to-one correspondence between elements of the two sets. To establish such a correspondence, we first computed the DCJ-indel distance between every inferred genome and every simulated genome. To reduce the resulting all-to-all comparisons to the best one-to-one correspondence, we interpret these comparisons as a complete bipartite graph where each partition corresponds to one of the sets of constituent genomes. We then can compute a minimum weight matching (MWM) to identify the set of genomes that minimize the total DCJ-indel distance. We can consider this matching to provide the most similar correspondence between the inferred and simulated sets. The DCJ-indel distances associated with the matched pairs can be used to assess the structural accuracy of the inferred pangenome. Lower distances indicate a more faithful reconstruction of the original genome organization.

\begin{figure}[h]
    \centering
    \includegraphics[width=0.7\textwidth]{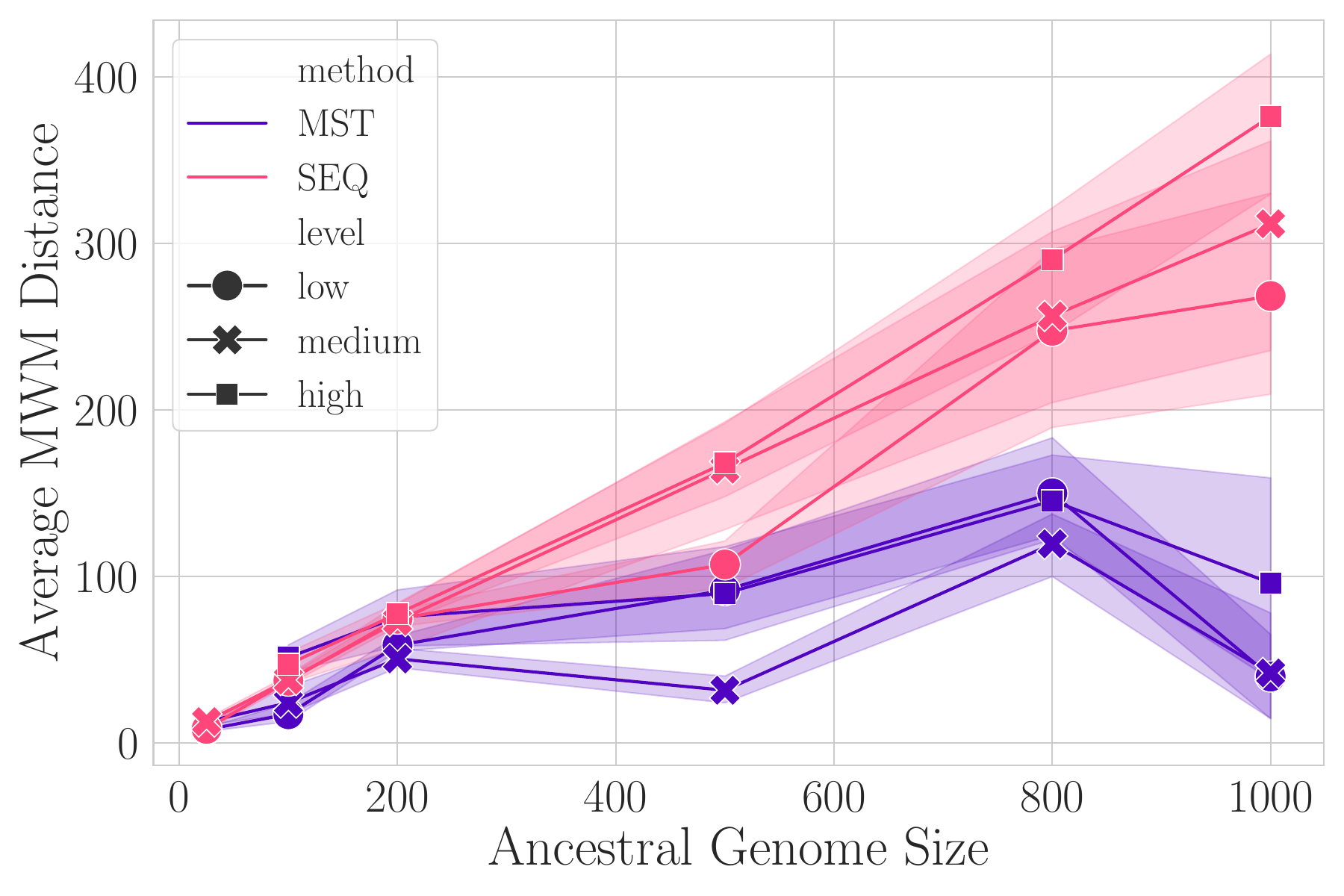}
    \caption{ Summary of average MWM DCJ-id distance among all the simulated datasets. Lower distances correspond to a more accurate reconstruction of the genome structure.}
    \label{fig:MWM_single_general}
\end{figure}

Figure~\ref{fig:MWM_single_general} illustrates the structural accuracy of the inferred pangenomes, summarizing the average matched DCJ-indel distance across all simulation settings. Overall, the MST method consistently yields lower average distances than SEQ indicating a more accurate recovery of the original genome. 

\section{Conclusion}

In this work we introduced the pangenome peeling problem and  two algorithms that construct pangenomes explaining a given weighted adjacency graph. 
We observed differences in the inferred number of constituent genomes, the recovered core gene set content and the structural similarity of the reconstructed genomes using the DCJ-indel distance. Across nearly all simulation settings, the maximum spanning tree method outperformed the sequential edge addition method, yielding more accurate estimates of pangenome size and structure.

Several directions for future work remain. First, the current model assumes that all genes are unique and does not account for gene duplications. Extending the framework to work with duplicated genes would increase its biological significance. Second, incorporating prior information, such as known or partially known core or shell markers, may help reduce the ambiguity of the peeling problem and perhaps improve reconstruction accuracy.

\bibliographystyle{unsrt}  
\bibliography{references} 

@article{jayakodi, author={ Jayakodi, Murukarthick and Shim, Hyeonah  and  Mascher, Martin}, year= {2025}, title= {What Are We Learning from Plant Pangenomes?},journal={ Annual Review of Plant Biology},volume= {76}, pages={663-686}, url= {https://doi.org/10.1146/annurev-arplant-090823-015358}
}

@article{secomandi2025pangenome,
  title={Pangenome graphs and their applications in biodiversity genomics},
  author={Secomandi, Simona and Gallo, Guido Roberto and Rossi, Riccardo and Rodr{\'\i}guez Fernandes, Carlos and Jarvis, Erich D and Bonisoli-Alquati, Andrea and Gianfranceschi, Luca and Formenti, Giulio},
  journal={Nature Genetics},
  volume={57},
  number={1},
  pages={13--26},
  year={2025},
  publisher={Nature Publishing Group US, New York}
}

@ARTICLE{liu,
    
AUTHOR={Liu, Chong  and Xu, Hang  and Li, Zheng  and Wang, Yukun  and Zhang, Jiaxian  and Qiao, Siwei  and Zhang, Hao },
           
TITLE={From map to blueprint: the plant pan-genome unraveling genetic mysteries and powering precision breeding},
          
JOURNAL={Frontiers in Plant Science},
          
VOLUME={Volume 16 - 2025},
  
YEAR={2025},
  
URL={https://www.frontiersin.org/journals/plant-science/articles/10.3389/fpls.2025.1673637}}

@article{li2014novo,
  title={De novo assembly of soybean wild relatives for pan-genome analysis of diversity and agronomic traits},
  author={Li, Ying-hui and Zhou, Guangyu and Ma, Jianxin and Jiang, Wenkai and Jin, Long-guo and Zhang, Zhouhao and Guo, Yong and Zhang, Jinbo and Sui, Yi and Zheng, Liangtao and others},
  journal={Nature biotechnology},
  volume={32},
  number={10},
  pages={1045--1052},
  year={2014},
  publisher={Nature Publishing Group US, New York}
}

@article{wang2023pangenome,
  title={A pangenome analysis pipeline provides insights into functional gene identification in rice},
  author={Wang, Jian and Yang, Wu and Zhang, Shaohong and Hu, Haifei and Yuan, Yuxuan and Dong, Jingfang and Chen, Luo and Ma, Yamei and Yang, Tifeng and Zhou, Lian and others},
  journal={Genome Biology},
  volume={24},
  number={1},
  pages={19},
  year={2023},
  publisher={Springer}
}

@article{lgtdead,
  title={Lateral gene transfer from the dead},
  author={Sz{\"o}ll{\H{o}}si, Gergely J and Tannier, Eric and Lartillot, Nicolas and Daubin, Vincent},
  journal={Systematic biology},
  volume={62},
  number={3},
  pages={386--397},
  year={2013},
  publisher={Oxford University Press}
}

@Article{eizenga,
  title={Pangenome graphs},
  author={Eizenga, Jordan M and Novak, Adam M and Sibbesen, Jonas A and Heumos, Simon and Ghaffaari, Ali and Hickey, Glenn and Chang, Xian and Seaman, Josiah D and Rounthwaite, Robin and Ebler, Jana and others},
  journal={Annual review of genomics and human genetics},
  volume={21},
  number={1},
  pages={139--162},
  year={2020},
  publisher={Annual Reviews}
}

@Article{bao,
author={Bao,Z and  Weigel, D.}, title= {Complexity welcome: Pangenome graphs for comprehensive population genomics}, 
journal= {Quantitative Plant Biology}, 
year= {2025}, 
volume={6}, 
pages={e43},
 \url={doi:10.1017/qpb.2025.10028}
}

@Article{odgi,
  title={\textsc{Odgi}: understanding pangenome graphs},
  author={Guarracino, Andrea and Heumos, Simon and Nahnsen, Sven and Prins, Pjotr and Garrison, Erik},
  journal={Bioinformatics},
  volume={38},
  number={13},
  pages={3319--3326},
  year={2022},
  publisher={Oxford University Press}
}

@article{akiyama1980covering,
  title={Covering and packing in graphs. III: Cyclic and acyclic invariants},
  author={Akiyama, Jin and Exoo, Geoffrey and Harary, Frank},
  journal={Mathematica Slovaca},
  volume={30},
  number={4},
  pages={405--417},
  year={1980},
  publisher={Mathematical Institute of the Slovak Academy of Sciences}
}

@article{nontrivialpc,
  title={Nontrivial path covers of graphs: existence, minimization and maximization},
  author={G{\'o}mez, Renzo and Wakabayashi, Yoshiko},
  journal={Journal of Combinatorial Optimization},
  volume={39},
  number={2},
  pages={437--456},
  year={2020},
  publisher={Springer}
}

@article{unimog,
  title={\textsc{UniMoG}—a unifying framework for genomic distance calculation and sorting based on DCJ},
  author={Hilker, Rolf and Sickinger, Corinna and Pedersen, Christian NS and Stoye, Jens},
  journal={Bioinformatics},
  volume={28},
  number={19},
  pages={2509--2511},
  year={2012},
  publisher={Oxford University Press}
}

@article{indels,
  title={Double cut and join with insertions and deletions},
  author={Braga, Mar{\'\i}lia DV and Willing, Eyla and Stoye, Jens},
  journal={Journal of Computational Biology},
  volume={18},
  number={9},
  pages={1167--1184},
  year={2011},
  publisher={SAGE Publications Sage CA: Los Angeles, CA}
}
\end{document}